\documentstyle[seceq]{ptptex}

\def\DM{Dzyaloshinsky-Moriya }

\title{
Dynamics, Selection Rules and Dzyaloshinsky-Moriya interactions
in Strongly Frustrated Magnets.

}

\author{
Olivier {\sc C\'epas}$^{a,b}$, Toru {\sc Sakai}$^{c}$ and  Timothy {\sc Ziman}$^{a}$ \footnote{Also at CNRS, LPM2C, UMR 5493, Grenoble} \footnote{E-mail addresses: cepas@physics.uq.edu.au,tsakai@cc.tmit.ac.jp,ziman@ill.fr} 
}

\inst{
$^a$Institut Laue Langevin, B.P. 156, 38042 Grenoble,France 
\\
$^b$ Department of Physics, University of Queensland, QLD 4072 Australia. 

$^{c}$Tokyo Metropolitan Institute of Technology, Asahigaoka, Hino, Tokyo 191-0065, Japan

}

\recdate{
}

\abst{
Anisotropic spin-spin interactions of the symmetry described
by  Dzyaloshinsky and Moriya are generally considered weak, as they depend
on the spin-orbit couplings. In frustrated spin systems with singlet
ground states they can, however, have rather strong effects. We discuss
recent results related to 
two gapped spin systems: $\rm CuGeO_3$ and $\rm SrCu_2(BO_3)_2$
in particular.
In the first compound the  Dzyaloshinsky-Moriya interactions effectively
lower the symmetry of the magnetic unit cell and this leads to
doubling of the low frequency mode.
In the second case, the Dzyaloshinsky-Moriya interactions
also split the lowest magnon mode linearly in the spin-orbit coupling. In addition, 
the relatively weak
Dzyaloshinsky-Moriya interactions can dominate the dispersion.

Consideration of the selection rules for optical transitions
show that while the  Dzyaloshinsky-Moriya interactions can
explain much of the  dynamics, they do 
not explain the observed transition amplitudes.
This leads to a review of 
recent calculations of anisotropic spin-phonon couplings.
We discuss how
this leads to a novel mechanism to explain the ESR intensities
in the spin gap systems discussed. Selection rules for this 
novel mechanism involving coupling to the electric field of the
resonant probe are discussed and relation to polarised neutron
experiments briefly mentioned. 

} 
\begin{document}

\maketitle

\section{Introduction}
In strongly frustrated magnets with singlet ground states
\DM interactions\cite{rf:Dzyaloshinski,rf:Moriya}
$\sum_{i,j}  \vec{D}_{i,j}
.(\vec{S}_{i} \times \vec{S}_{j})$, (with sum over  neighbours $i$ and $j$),
can have marked effects on  the dynamics even though they are generally
considered to be a relatively weak perturbation of the isotropic
exchange. 
By strongly frustrated we mean systems that have singlet
ground states separated by a gap. This may be associated with 
a  spin-Peierls distortion, as exemplified by the compound
$\rm CuGeO_3$ 
or  purely geometric frustration, as in the case of the 
compound $\rm SrCu_2(BO_3)_2$ which is close to a model system for  the Shastry-Sutherland
model in two dimensions. 

Moriya\cite{rf:Moriya} estimated that the magnitude of the \DM vector
between two sites is related to the isotropic exchange $J$  by the relation 
 $D = (\frac{\Delta g}{g}) J$ if it is allowed by symmetry, where
$\Delta g = g-2$ is the measure of the strength of spin-orbit
interactions, about .1 in the case of the copper oxides.
$J$ is the isotropic Heisenberg exchange. We remark that one can
find exceptions to the rule that the \DM interaction should
be much smaller than the isotropic exchange in model calculations, essentially
by involving superexchange with copper-oxygen-copper angles close to $\frac{\pi}{2}$ , in which case $J$ is exceptionally small.
These exceptions involve fine-tuning and have, as yet, not been shown
to be relevant to real systems. 
Local symmetries or approximate symmetries may of course
give a \DM coupling that is much smaller: if the relevant exchange $J$ comes
from  superexchange paths that have an  inversion symmetry $\vec{D}$ must vanish, and if this inversion symmetry is only
weakly broken  $\vec{D}$ must be small.
\par
We shall argue nonetheless that $\vec D$ can dominate certain
features of the dynamics because:
\newline
(i)\ it is the leading source of spin anisotropy in zero field, and
\par\noindent
(ii)\ it may lower the spatial symmetry of the effective magnetic model.
\newline
In each case it may be expected to allow transitions forbidden
by the original spin or space symmetries. Selection rules
are necessary  to determine experimentally the $\vec D$ vectors
and see what processes are allowed and distinguish from the 
effects of other anisotropies, for example
staggered {\boldmath g} tensors in finite magnetic field. In some cases the \DM 
interaction alone
does not permit  transitions that have 
actually been observed, and this will lead us to consider a higher order
of anisotropy: ``dynamical Dzyaloshinsky-Moriya'', in which the spin anisotropy is generated
by distortions of the equilibrium lattice linear in the phonon
coordinates. By a perturbative treatment of this coupling
we derive an effective operator purely in terms
of spin-operators and again give selection rules.  These
may explain optical transitions, at wave vectors $q = 0$, observed by ESR and infrared absorption,
and, for finite values of $q$, mixing of nuclear and magnetic neutron scattering amplitudes. 
\par
We remark that in the frustrated cases we are discussing, effects such as
the splittings
may appear {\it linearly} in the strength of the \DM coupling: this is in contrast
to the case of ordered antiferromagnets where, for example  the 
contribution to the energy of 
a weakly ferromagnetic state is quadratic in the spin-orbit strength.
In that case the 
exchange anisotropy, which is also quadratic \cite{rf:Moriya},
may compensate at least in special cases\cite{rf:Kaplan,rf:Aharony}.
Here the exchange anisotropy is of higher order and can safely be neglected.

\par
This paper will review material presented in greater detail elsewhere, either for the 
static \DM  \cite{rf:SakaiCZ} and the dynamic \cite{rf:these,rf:OlTim}.

\section{Dynamics: Examples of the influence of \DM }
In this section we will discuss two cases where the direction of the \DM vectors
can be predicted from the knowledge of the structure and have marked
effects on the dynamics, producing  an effective doubling or tripling
of the low frequency mode, as observed in inelastic scattering
of neutrons or in absorption of  light.     

\subsection{$\rm CuGeO_3$}
We first consider the case of $\rm CuGeO_3$, that has been much studied
as the first inorganic example of a spin-Peierls system. In fact
analysis of the magnetic susceptiblity has shown that this system
is, in addition to being a spin-Peierls system, magnetically
frustrated.  
It may be described by the Hamiltonian
\begin{eqnarray}
{\cal H} &=& {\cal H}_{1D} + {\cal H}_t \\ {\cal H}_{1D} &=&
\sum_{i,j} J_c (1+\delta(-1)^{i+j})\vec{S}_{i,j}.\vec{S}_{i+1,j} +
J_{2c} \vec{S}_{i,j}.\vec{S}_{i+2,j} \\ {\cal H}_t &=& \sum_{i,j}
J_b \vec{S}_{i,j}.\vec{S}_{i,j+1}
\label{hamiltonien}
\end{eqnarray}
$c$ and $b$ refer to crystalline axes of strongest and next-to-strongest
magnetic exchange.    
The argument $i$ is in the chain direction $c$, and $j$ in the 
transverse direction $b$.
The low energy magnetic excitations are
well described by an alternating exchange $J_c(1 \pm  \delta)$ and second-nearest-neighbour
coupling $J_{2c}$ and an interchain coupling $J_b$. The numerical
values of the couplings can be estimated, including  the effects of interchain
coupling $J_b/J_c=0.15$,  as $J_{2c}/J_c = 0.2$, 
dimerization $\delta \approx .065$, $J_c=12.2\ {\rm meV}$ \cite{rf:Bouzerar1}.
 The observation of a second
mode \cite{rf:Lorenzo,rf:Nojiri}(called an ``optical mode'' by the experimentalists), with weak
intensity, was initially attributed to a slight difference in the dimerization
of alternate chains\cite{rf:Bouzerar1} 
but in fact is more convincingly attributed to the \DM interactions.
The alternation of dimerization from chain to chain (the factor $(-1)^{i+j}$)
to give a chequer-board structure, is responsible for the fact that the
mode is out of phase with the stronger mode. 
From the observed structure\cite{rf:these} the \DM vectors should be in 
the $\vec{c}$ direction, act between spins in the perpendicular ($b$) direction and alternate:
 
\begin{eqnarray}
{\cal H}_{\perp}^{DM}=\sum_{i,j} (-1)^j D_{b}
\vec{c}.(\vec{S}_{i,j} \times \vec{S}_{i,j+1})
\label{hamiltonien anisotrope}
\end{eqnarray}
By making a rotation of the spin axes about the plane perpendicular
to the \DM vectors, following an argument of  Affleck and Oshikawa   
\cite{rf:Oshikawa,rf:OlTim}, the magnetic response  can be deduced from that without the \DM interactions.

\begin{eqnarray}
S^{aa}_{D}(\vec{q}, \omega) &=& \cos^2(\frac{\theta}{2}) S^{aa}(\vec{q},
\omega) + \sin^2(\frac{\theta}{2}) S^{bb}(\vec{q} - \vec{\pi}, \omega)
\nonumber \\ S^{bb}_{D}(\vec{q}, \omega) &=& \cos^2(\frac{\theta}{2})
S^{bb}(\vec{q}, \omega) + \sin^2(\frac{\theta}{2}) S^{aa}(\vec{q} - \vec{\pi} ,
\omega) \nonumber \\ S^{cc}_{D}(\vec{q}, \omega) &=& S^{cc}(\vec{q},
\omega)
\label{relations}
\end{eqnarray}
where $\theta$ is  given by $\tan
\theta=D_{b}/J_b$. $S^{\alpha\alpha}(\vec{q}, \omega)$ are the dynamical
structure factors for an isotropic model. $\vec{\pi} = (0,\pi)$
with respect to axes $(q_c,q_b)$ and we neglect dispersion in the
$a$ direction as it is very weak. There should also be  a weak exchange
anisotropy producing an unobservably small splitting of the ``acoustic mode''
but we shall neglect this.
Thus the ``optic mode'' is in fact the
same mode seen at a different momentum transfer and should be visible with 
relative intensity: $({\frac{D_b}{J_b}})^2$.  
 From the observed intensity\cite{rf:Lorenzo}, this gives an 
estimate of the magnitude of the \DM vector as $D \approx 0.4 $meV.
We remark that a test of this mechanism should be the behaviour
in finite field: as only the two polarizations transverse to the 
direction of the \DM vector are involved in the doubling, in external
magnetic field parallel to $\vec{D}$ the ``optic mode'' should split into two branches 
\subsection{$\rm SrCu_2(BO_3)_2$}
The second case is that of Strontium Copper Borate. This 
compound
is very interesting in that restricting first to isotropic interactions,
it can be considered
as planes of spins $\frac{1}{2}$ interacting via 
the Hamiltonian of the Shastry-Sutherland model in two dimensions. This model
has the peculiarity that the product of singlet
states on the closest dimers with the stronger exchange $J$ is still an {\it exact} eigenvector when the frustrated second nearest neighbour interactions $J^{\prime}$ 
are included\cite{rf:ShastryS}. Furthermore this eigenvector is the ground state 
even for the relatively large value
of the relative coupling $J^{\prime}/J = 0.62$. 
This ratio is estimated either from the susceptibility\cite{rf:Miyahara} or the 
ratio of the energies of singlet states, seen in Raman scattering
to triplet energies, seen by magnetic neutron scattering\cite{rf:borate}.
The  interaction between planes
is via couplings that are both weak and frustrated. When we take into account anisotropies the
ground state will be perturbed. Nevertheless we  have  
a rare example of a system with exponentially decaying magnetic correlations
and a   ground state that can be  described as a local product of dimers with small corrections.
Here we shall in fact
consider a slightly idealized view of the compound, ignoring a small buckling
of the planes. In this case the \DM couplings are strictly perpendicular
to the planes, act between the next-nearest neighbour copper
ions and, as shown in \cite{rf:borate}, give fine structure to 
the lowest lying magnon: i.e. a small splitting into three modes, as had been
observed in 
the optical experiments of Nojiri et al\cite{rf:Nojiri2} and the neutron
inelastic scattering. Taking into account renormalisation
of the gap by the frustrated interactions, the splitting can be used to derive a precise
numerical value of the \DM vectors $\vec{D}^c = 0.18\  {\rm meV}$. 
\par
The first  effect
of the \DM interactions  is then  to split the
original triplet states. The more striking effect is that this splitting
can dominate the dispersion.
Propagation of the magnons in the Shastry-Sutherland lattice is weak: frustration of the interdimer
couplings leads to a bandwidth that begins in sixth order in $J^{\prime}/J$.
The \DM interaction, in contrast, is {\it not} frustrated and the splitting is
linear in $|\vec{D}|$. Thus the splitting due to \DM interaction is
estimated to be larger than that of the dispersion due to interdimer coupling, even though
that coupling is much larger\cite{rf:borate}.    
\section{Selection rules for \DM interaction}
In the optical experiments of Nojiri et al\cite{rf:Nojiri,rf:Nojiri2}, the resonance is from the 
ground state to the excited magnetic states. The observation of absorption
requires some anisotropies: as the ground state without anisotropies
is a spin singlet the operator corresponding to  coupling with the probe magnetic
field $\vec{h}.\sum_{i} S_i$ applied to the ground state  vanishes. As the \DM interaction
mixes in non-singlet components the matrix elements to  excited states
may be non-zero. We can first estimate  
which ones are non-zero, and the dependence of the absorption strengths on external field,
if we consider strictly local symmetries\cite{rf:Kokado,rf:SakaiCZ}. This means 
we considering two spins in external uniform applied field $\vec{H}$ and with
different possible  polarizations
of the resonating probe field $\vec{h}$:
\begin{eqnarray}
{\cal H}{({\vec S}_1,{\vec S}_2)} = J{\vec S}_1\cdot {\vec S}_2+
{\vec D}\cdot ({\vec S}_1\times {\vec S}_2) 
-{\vec H}\cdot ({\vec S}_1+{\vec S}_2)-{\vec h}(t)\cdot ({\vec S}_1+{\vec S}_2), 
\end{eqnarray}
where ${\vec D}$ is the \DM vector 
and ${\vec H}$ is 
the external magnetic field. If $\vec H$ is parallel to ${\vec D}$
then the component of total spin in their common direction ($z$ let us say)
is a constant of the motion. Therefore only polarisations of $\vec h$ perpendicular to  $\vec H$ or ${\vec D}$
will give absorption to states with $\Delta S^z = {\pm} 1 $. The strength of absorption will be independent of 
the field as the eigenvectors do not change with  $H$.
For $\vec H$ perpendicular to  ${\vec D}$, the total spin along
the axis of   ${\vec D}$ is no longer a constant of the motion: therefore
there will be field dependence of the absorption of the three
different components of the resonating field. These general properties
are clearly shared by the lattice model but the exact dependencies
for  $\vec H$ perpendicular to  ${\vec D}$ must be calculated.
Explicit results are given in  
reference \cite{rf:SakaiCZ}. Such selection rules are  used to verify the direction of the 
\DM vector, especially in the case when symmetry alone cannot uniquely
determine its direction.  In addition there are further constraints we can
call ``lattice selection rules'' which depend on the overall
pattern of \DM vectors. Applied to the two structures we are considering,
these have interesting consequences: in the case of 
$\rm CuGeO_3$, from the argument we have mentioned of a rotation
of axes of the spin variables, only the ``optic mode'' should be visible at $q = 0$.
This was in agreement with older results\cite{rf:Boucher,rf:Loosdrecht}, but the recent results of 
Nojiri et al\cite{rf:Nojiri} showed that both modes were visible. Similarly in the case
of the {$\rm SrCu_2(BO_3)_2$}, a lattice symmetry (reflection in a diagonal
followed by rotation by $\pi$) leads to a zero amplitude for excitation
of the triplet states, even in the presence of the 
\DM couplings. Of course, there are additional
anisotropies due to slight buckling of the planes and anisotropies of the 
{\boldmath g } tensors, but nevertheless the amplitude of the absorption 
in the two cases is somewhat surprising and this leads us to consider an
alternative explanation in terms of a dynamical \DM interaction. 
\section{Dynamical \DM interaction }
We shall now consider a general anisotropic
spin-phonon couplings corresponding to modulation of the 
exchange by linear coupling to lattice distortions. The 
term in the Hamiltonian coupling the phonon and spin operators is:

\begin{equation}
{\cal H}^{\prime} = \sum_{ijd\alpha\beta} g_{d}^{\alpha} u^{\alpha}_{id}
\vec{S}_i.\vec{S}_{j} + d^{\alpha \beta}_d u^{\alpha}_{id}
(\vec{S}_i \times \vec{S}_{j+1})^{\beta} 
\label{spin-phonon coupling}
\end{equation}
where $u^{\alpha}_{id}$ is the $\alpha$ component of the displacement
operator of atom $d$ in unit cell $i$, $g_d^{\alpha}$ and $d^{\alpha
\beta}_d$
are, respectively, the isotropic spin-phonon coupling and the
dynamical Dzyaloshinski-Moriya interaction.

A typical case in Copper oxide is that there are frequently bridges
of Cu$_2$O$_2$ with inversion symmetry in the equilibrium state. 
In the presence of a phonon, the atomic positions may move so as to 
instantaneously remove the inversion symmetry, generating a 
\DM anisotropy.

Consideration of ``dynamical'' \DM terms were in fact motivated
first by experiments in inelastic neutron scattering in which
by measurement of the polarisation of 
scattered neutrons  one can probe mixed ``nuclear'', i.e. involving
the positions of the nuclei that scatter from neutrons via the strong interaction
and ``magnetic'', i.e. interactions from the magnetic fields generated
by the spin and orbital moments of electrons\cite{rf:Blume,rf:Maleyev_neut}
. In this case the geometry
of the experiment is such that only correlations between the two
terms can give a non-zero results, and furthermore, as a rotation can
be measured only an interaction with a ``handedness'' such as the \DM
interaction can give a non-zero result\cite{rf:Maleev_pol}.  
 
\subsection{Consequences for optical experiments: $\vec{e}$ field absorption}
In this paper
we will not consider the effect in neutron scattering ( see reference\cite{rf:OlTim}) but the analogous effect in optical absorption. The essential
point is that as the spin-orbit interaction mixes orbital
and spin degrees of freedom, the separation  between coupling
to the magnetic field and the electric fields of the probe is 
no longer complete: in fact the electric field, which one would
expect to couple only to the dipole electric moments
will also effectively couple to the spins and give absorption
to excitations considered normally simply ``magnetic''. The effects can
be calculated perturbatively in the spirit of Fleury and Loudon\cite{rf:Loudon} for 
Raman absorption to magnetically excited states, but with the difference
that the spin-orbit interaction is included, and that the excited states
are involving a phonon excitation rather than an electronic excitation\cite{rf:Lorenzana-S}.
The linear Hamiltonian is applied to the ground state and the 
excited magnetic state of the unperturbed Hamiltonian to first order.
The  matrix element of the electric dipole operator  between the perturbed states $0^{\prime}$ and $\alpha{\prime}$ including $\cal{H}^{\prime}$ can then be written as that of an effective
operator acting between the unperturbed states $0^{}$ and $\alpha{}$. This
operator is purely written in terms of spin operators:

\begin{eqnarray}
\langle \alpha^{\prime} | \sum_{id} q_d \vec{u}_{id}. \vec{e}   | 0^{\prime} \rangle &=& 
\langle \alpha \mid \sum_{ij} \gamma \vec{S}_i . \vec{S}_{j} + \vec{\delta}.(\vec{S}_i \times \vec{S}_{j}) \mid 0 \rangle \label{result} \\
\gamma &=&  \sum_s \frac{ \Omega_s}{\omega_{\alpha}^2 - \Omega_s^2} g_s 
 (\vec{\cal D}_s. \vec{e}) \\
\vec{\delta} &=&  \sum_s \frac{ \Omega_s}{\omega_{\alpha}^2 - \Omega_s^2} \vec{d}_s  (\vec{\cal D}_s. \vec{e})
\end{eqnarray}
 
\noindent 
where $\vec{\cal D}_s= \sum_d q_d \vec{\lambda}_{ds(q=0)}$ is the amplitude
of the instantaneous electric dipole of the unit cell due to the phonon
mode $s$ with energy $\Omega_s=\Omega_{(q=0,s)}$ which displaces the charges
$q_d$ . The final magnetic
state has an energy $\omega_{\alpha}$. $g_s=\sum_{d,\alpha} g_d^{\alpha}
\lambda^{\alpha}_{ds}$ is the amplitude of the variation of the
magnetic exchange energy due the atomic distortions of the phonon $s$
($\lambda^{\alpha}_{ds}$ is the amplitude of the motion of the atom
$d$, in the direction $\alpha$ due to the phonon $s$ at $q=0$).
Here the sum $ij$ is assumed to run over a set of equivalent neighbours:
more generally there could be a set of $\gamma$ and $\delta$ for different
inequivalent neighbours.
The selection rules for the contribution of a particular phonon mode $s$
to contribute are that: 

\begin{itemize}
\item (i)
$(\vec{\cal D}_s.\vec{e}) \neq 0$: the virtual phonon $s$ creates distortions that carry an instantaneous electric dipole ${\cal D}_s$. In other words, the phonon $s$ must be optically active.
\item (ii) \begin{itemize}
\item $g_s \neq 0$: The distortion of the unit cell due to the phonon
$s$ modulates the magnetic exchange between the spins. Only  spin-conserving transitions
at $\Delta S_{tot}=0$ are allowed.
\item $\vec{d}_s \neq 0$:  The distortion of the unit cell due to the phonon $s$ must break instantaneously the symmetry by inversion at the middle of the bond; so as to allow an instantaneous \DM interaction of amplitude $\vec{d}_s$. 
Transitions between different spin states $\Delta S_{tot} =0,\pm 1$ are allowed.
\end{itemize}
\end{itemize}
Note that the selection rules involve detailed knowledge of different phonons. 
Directions of the vector  $\vec{d}_s$ 
are constrained by the symmetry rules for static \DM
interactions applied to  structure distorted by the given phonon $s$
from the equilibrium structure.
The factors
$(\vec{\cal D}_s.\vec{e})$ can be measured independently
from the intensity at the frequency $\Omega_s$ of the {\it real} 
phonon creation.
For external magnetic field parallel to $\vec{\delta}$ the total
component of spin in this common direction $\alpha$ say  is conserved ( if this is the sole form of anisotropy or, if not, if this direction is  an axis of symmetry
shared with  the other anisotropies) and only transitions to the field-independent
level $\Delta S^{\alpha} = 0$ should be observed: thus the selection rule is quite different
from that for magnetic transitions with a (static) \DM interaction in the
same direction. Again for a field in the transverse directions there
will be transitions to the three states with magnetic field dependence that
could be calculated as in reference\cite{rf:SakaiCZ}. Note that
the wave functions must include any static \DM interaction $\vec{D}$  and the 
matrix elements involve in general different vectors  $\vec{\delta}$.
From the absolute intensities one should be able to deduce the
magneto-elastic constants $g_s$, and the components of ${\vec d}_s$.
In a  full comparison to 
experiment  it is desirable to control  the polarisations
of the $\vec{e}$ and $\vec{h}$ fields of the probe  separately.
Frequently  only the direction of  propagation,  i.e. their vector product,
is controlled with respect to the crystal axes. Recent experiments
by R$\tilde{\mbox{o}}$$\tilde{\mbox{o}}$m et al.\cite{rf:Room}
of infrared absorption with  polarised electromagnetic waves
seem to be consistent with the selection rules enunciated: for example
in  $\rm CuGeO_3$ extinction for $\vec{e} \parallel c $ \cite{rf:Damascelli,rf:Room-cugeo} follows
as the  mirror planes of the equilibrium structure
are maintained under an assumed distortion of the atoms
along the $c-$axis. In {$\rm SrCu_2(BO_3)_2$} we have also found\cite{rf:OlTim}
good agreement with the experiments\cite{rf:Room}, at least 
by using a simplified view of the structure. If, for example,
$\vec{e}$ is taken in the $(ab)$ plane and we assume that even with
the virtual phonon that couples to such an electric field  the 
 $(ab)$ plane remains a 
mirror plane. In this case the effective operator, by the standard symmetry arguments, 
will have components along the $c$-axis only. As argued above, there should be absorption to the $S_z = 0$ mode only, provided the external magnetic field
$\vec{H}\parallel c$, and field-dependent absorption to the
(static) \DM interaction split lines for $\vec{H} \perp c$.

We will not  compare to the  neutron case in detail\cite{rf:these,rf:OlTim} butnote that
in calculating the relevant matrix element for ``nuclear'' scattering
to a magnetic state,  while
the same magneto-elastic constants and vectors will enter,  the  vectors $\vec{\delta}$
will differ as $(\vec{\cal D}_s.\vec{e})$ for example will be replaced
by the phonon form factor for nuclear neutron scattering. The selection 
rules involve the scattering geometry, and therefore different phonons may contribute.

\section{Conclusions}
We have reviewed results for the selection rules governing
optical absorption, in particular in the presence of both
static \DM interactions and terms generated by coupling to phonons
that lower the symmetry. In the second case both nuclear and magnetic
scattering amplitudes are mixed in inelastic neutron scattering, and optically, magnetic states may be excited
by the electric field component of the probe. Testing of these effects can be by a full polarization
experiments in both cases:  in  neutron  scattering by polarisation of both
incoming and outcoming beams, and, in the optical experiments,  by controlling the
polarisation of the electric and magnetic components.

\section*{Acknowledgements}
We would like to thank H. Nojiri for many communications and discussion of 
unpublished results and J.P. Boucher
for constant encouragement and debate.

\appendix


\end{document}